\documentclass[3p,times]{elsarticle}

\usepackage{ecrc}
\usepackage{horowitzarXiv}
\usepackage{color}
\usepackage{hyperref}
\definecolor{darkgreen}{rgb}{0,.7,0}

\hypersetup{
    unicode=true,          
    pdftitle={Qualitative and Quantitative Energy Loss?},    
    pdfauthor={W. A. Horowitz},     
    pdfnewwindow=true,      
    bookmarksnumbered=true,
    colorlinks=true,       
    linkcolor=red,          
    citecolor=darkgreen,        
    filecolor=magenta,      
    urlcolor=cyan,           
    menucolor=blue,
    baseurl=http://www.arxiv.org/abs/,
    linktocpage=true
}

\volume{00}

\firstpage{1}

\journalname{Nuclear Physics A}

\runauth{W.\ A.\ Horowitz}

\jid{nupha}

\usepackage{amssymb}

\usepackage[figuresright]{rotating}

\begin{document}

\begin{frontmatter}



\dochead{}

\title{Qualitative \emph{and} Quantitative Energy Loss?}

 \author[label1,label2]{W.\ A.\ Horowitz}

 \ead{wa.horowitz@phy.uct.ac.za}
 \ead[url]{www.phy.uct.ac.za/people/horowitz}

 \address[label1]{Department of Physics, University of Cape Town, Private Bag X3, Rondebosch 7701, South Africa}
 \address[label2]{Department of Physics, The Ohio State University, 191 West Woodruff Avenue, Columbus, OH, 43210, USA}



\begin{abstract}
Despite the 300\% systematic theoretical uncertainty associated with extracting \qhat using current energy loss models due to the collinear approximation, and despite the uncertainties in the initial geometry and corrections due to the fluctuations in the initial overlap of a heavy ion collision, a simultaneous description of \highpt \pizero \raa and \vtwo is not possible with current pQCD-based energy loss models.  However a good description of out-of-plane \raa as a function of centrality is possible.  Alternatively, energy loss models based on AdS/CFT give a better qualitative description of \pizero and non-photonic electron \raa as well as \highpt ratios of $\bar{p}$ to $\pi^-$ production than those based on pQCD.
\end{abstract}

\begin{keyword} quark-gluon plasma \sep pQCD \sep AdS/CFT \sep energy loss


\end{keyword}

\end{frontmatter}


\section{Introduction}\vspace{-.05in}
High momentum (\highptcomma) particles are ideally suited for probing the properties of the quark-gluon plasma (QGP) created in heavy ion collisions \cite{Wiedemann:2009sh,Majumder:2010qh}.  In particular they provide an independent test of whether the medium is best described by traditional weakly-coupled perturbative QCD (pQCD) or the more novel strongly-coupled methods of anti-de Sitter/conformal field theory (AdS/CFT) \cite{Horowitz:2010dm}.

In the early years of physics at the Relativistic Heavy Ion Collider (RHIC) energy loss models based on pQCD had seemingly quantitative success in describing the suppression of \highpt particles, in particular the $R_{AA}^\pi(p_T)$ \cite{Akiba:2005bs}.  However soon afterwards it was found that, taken as a broad class, these pQCD-based models do not do a good job of even qualitatively describing several other observables (while simultaneously maintaining the description of the normalization of the pion suppression), for instance: the azimuthal anisotropy of the pion suppression pattern as measured by the pattern's second Fourier coefficient \cite{Shuryak:2001me,Adare:2010sp}, $v_2$; the \raa and \vtwo of non-photonic electrons \cite{Wicks:2005gt,Adare:2006nq}; or the back-to-back suppression of particles as measured by $I_{AA}$ \cite{Adams:2006yt,Armesto:2009zi}.  

Several of these observables had their shortcomings, mostly due to a limited measured range in momentum.  For instance the current generation of energy loss calculations for heavy quarks assume that the quantity $M_Q/p_T$ is small, which is not particularly the case for the parent bottom quark of an observed 8 GeV non-photonic electron \cite{Djordjevic:2004nq}.  Similarly the pion $v_2$ was also measured only to $\sim8$ GeV, in which case one might claim that non-perturbative effects, for instance from hadronization \cite{Horowitz:2005ja}, might play an important role.  Being somewhat provocative, though, the recent measurements \cite{Wei:2009mj,Adare:2010sp} of pion \vtwo out to $\sim14$ GeV could be thought of as the nail in the coffin of pQCD-based energy loss models; it is difficult to reconcile the large ($\sim2$) discrepancy between the theoretical calculations and the data.  Before making such strong conclusions perhaps we should more fully explore the possible sources of uncertainty and missing physics in the current set of energy loss calculations, which has so far been somewhat limited \cite{Horowitz:2009eb}.

\vspace{-.05in}
\section{Uncertainties and Missing Physics in pQCD Energy Loss}
\vspace{-.05in}
There are many possible sources of uncertainties and missing physics in pQCD-based energy loss calculations.  In particular it is quite likely that there are large corrections due to: running coupling \cite{Horowitz:2006ya,Wicks:2007am}; a better treatment of the collinear approximation \cite{Horowitz:2009eb}; a better treatment of the parent quark mass; higher orders in opacity \cite{Gyulassy:2001nm,Wicks:2008ta}; a better treatment of elastic energy loss \cite{Wicks:2008}; and energy loss in cold, confined matter \cite{Domdey:2010id}.  We will show that the effect on \vtwo of varying the initial geometry is small.  Crucially, though, we find that we can describe very well the centrality dependence of the out-of-plane \raa suppression but not that of the in-plane \raacomma.  Perhaps the large discrepancy between theory and experiment in \vtwo is due to a missing coupling of the energy loss to flow.  

\begin{figure}
\begin{center}
$\begin{array}{cc}
\includegraphics[width=0.48\textwidth]{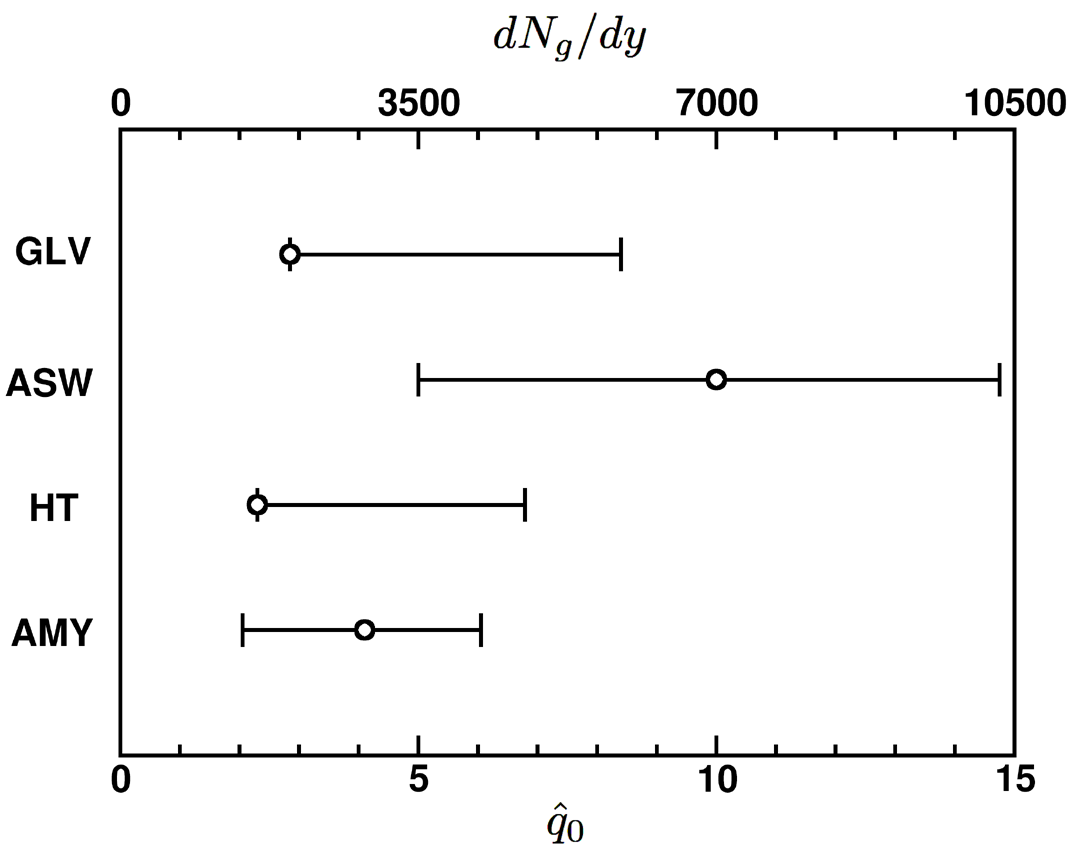}
\includegraphics[width=0.48\textwidth]{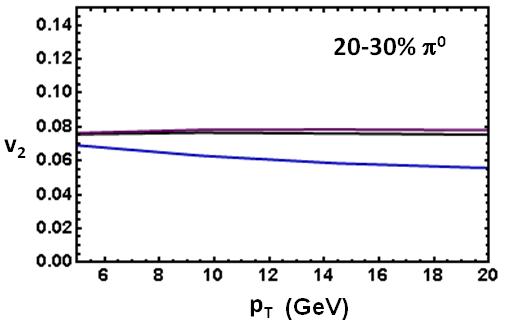}
\end{array}$
\end{center}
\vspace{-.1in}
\caption{\label{fig1}(Left) From Horowitz and Cole \cite{Horowitz:2009eb}, the uncertainty in the extracted \qhat due to the collinear approximation. (Right) The different \vtwo curves when the maximum opening angle for gluon radiation is varied as in \cite{Horowitz:2009eb} and the overall suppression, \raacomma, is fixed to data; note that there is little difference in the resulting \vtwo values. (See text for details.)}
\end{figure}

One sees in the left panel of \fig{fig1}, taken from \cite{Horowitz:2009eb}, an estimate of the uncertainty on the extracted \qhat from the various gluon bremsstrahlung energy loss formalisms due to the collinear approximation.  All current energy loss formalisms take only the lowest order term in $k_T/xE$, where $k_T$ is the momentum component of the radiated gluon perpendicular to the parent parton's direction of motion, $x$ is the fraction of momentum taken from the parent parton by the radiated gluon, and $E$ is the energy of the parent parton.  The uncertainty estimate comes, essentially, from allowing the maximum angle of emission for the radiated gluon to vary from 45$^\circ$ to 90$^\circ$.  This large uncertainty would seem to imply that there is little hope of falsifying any of these pQCD energy loss models.  However, one can drastically reduce this uncertainty when attempting to \emph{simultaneously} describe multiple observables.  In particular, the right panel of \fig{fig1} shows the result when one picks a maximum angle of emission, then varies the medium density in order to describe the overall pion suppression, \raacomma, and then one computes the resulting \vtwocomma; as one can readily see, there is little difference between the curves shown.  Therefore the very \highpt \vtwo puzzle cannot be resolved by claims of large theoretical uncertainty, at least as far as the collinear approximation is concerned.

\begin{figure}
\begin{center}
$\begin{array}{cc}
\includegraphics[width=0.48\textwidth]{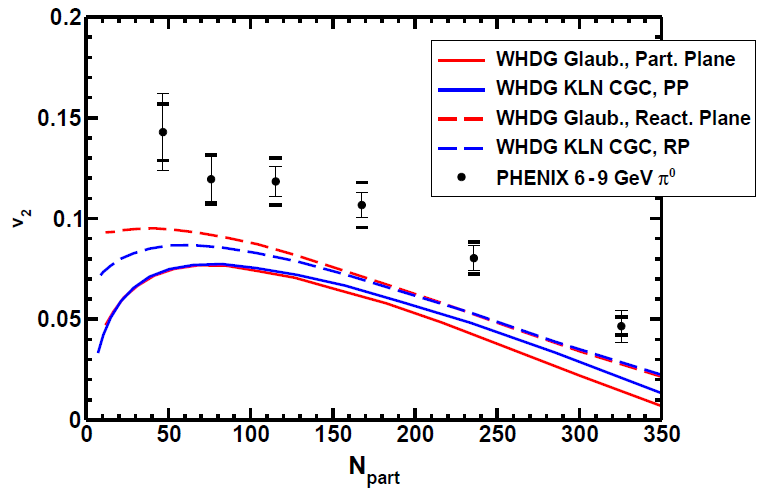}
\includegraphics[width=0.48\textwidth]{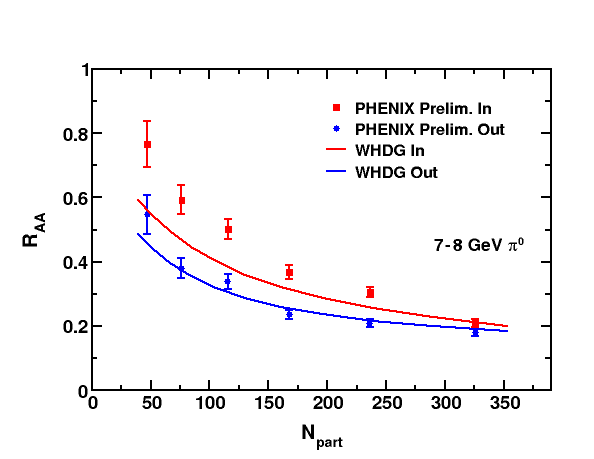}
\end{array}$
\end{center}
\vspace{-.1in}
\caption{\label{fig2}(Left) \vtwo as a function of \npartcomma.  While \vtwo is increased when taking into account the difference between the theoretical reaction plane and the measured participant plane and also the possibility of sharper edges of the medium due to Color Glass Condensate effects--while maintaining consistency with \raacomma--the theoretical values are still well below the experimental observations \cite{Wei:2009mj,Adare:2010sp}.  (Right) Surprisingly good agreement between WHDG and measurement for \raa out-of-plane as a function of centrality, while the disagreement in-plane is very large; data from \cite{Wei:2009mj,Adare:2010sp}.  Perhaps there are large coupling effects between energy loss and flow, minimized in the out-of-plane results, that show up as a large disagreement with \vtwocomma?}
\end{figure}

However another possible resolution to the \highpt \vtwo puzzle might come from the uncertainty due to the uncertainty in the initial conditions in a heavy ion collision.  The left panel of \fig{fig2} shows some exploration of the uncertainty due to collision geometry.  First, there is a possible correction to \vtwo from sharper or broader initial medium densities; we compare the results from a standard Glauber treatment \cite{Horowitz:2010dm} and one using the KLN model of the CGC \cite{Kharzeev:2001yq,Drescher:2006pi}.  Second there are corrections due to the fluctuations in initial conditions.  The work in \cite{Jia:2010ee} suggests that the largest effect of fluctuations in initial conditions comes from the difference between the theoretical reaction plane and the measured participant plane.  We compare the \vtwo results when calculations are done in the reaction plane or the ``rotated'' participant plane.  One can see from the left panel of \fig{fig2} that the uncertainty in \vtwo due to geometry considerations is not particularly large.  In particular, the maximum reasonable \vtwo calculation--one in which the difference between the theoretical reaction plane and measured participant plane is taken into account, the initial geometry is given by the KLN model of the CGC, and the 1-$\sigma$ upper bound of the medium density set by a comparison \cite{Adare:2008cg} with \raa is used--still significantly underpredicts the PHENIX \highpt \vtwo data.  

But what if we compare to a more differential measurement?  In the right panel of \fig{fig2} we show both the WHDG convolved radiative and elastic energy loss calculation of \raa in-plane and out-of-plane.  The agreement between the out-of-plane results to data as a function of centrality is, compared to the \vtwo results, amazingly good.  Perhaps, then, the disagreement between theory and experiment in the in-plane (and also, therefore the \vtwocomma) is due to \emph{missing} physics; since one expects the main physical difference from in-plane to out-of-plane to be from the difference in the underlying flow (more flow in-plane than out), perhaps the discrepancy is due to not including a coupling of energy loss to flow.  We should note that the effects \cite{Renk:2006sx} of coupling \cite{Baier:2006pt,Liu:2006he} flow to BDMPS energy loss \cite{Baier:1996kr,Baier:1996sk} appear small.  
Perhaps a better treatment of the coupling of GLV energy loss \cite{Gyulassy:2000fs,Gyulassy:2000er} to flow will yield theoretical calculations in agreement with the measured \vtwocomma.  Note that while for a fixed centrality there is a large difference between the pathlengths traveled in-plane vs.\ out-of-plane, the agreement for the out-of-plane results shown in the right panel of \fig{fig2} is over a range of centralities, and therefore, over a range of pathlengths.  Note also that \fig{fig2} demonstrates the extra utility gained by measuring \raaphi and not just \vtwocomma.

\vspace{-.05in}
\section{AdS/CFT}
\begin{figure}
\begin{center}
$\begin{array}{cc}
\includegraphics[width=0.48\textwidth]{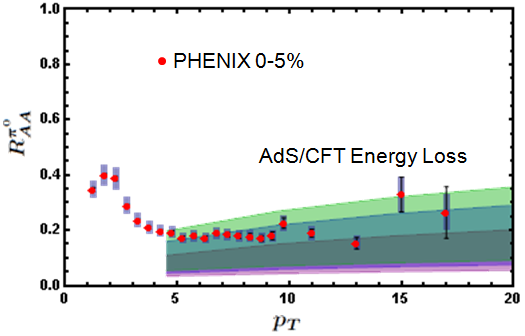}
\includegraphics[width=0.48\textwidth]{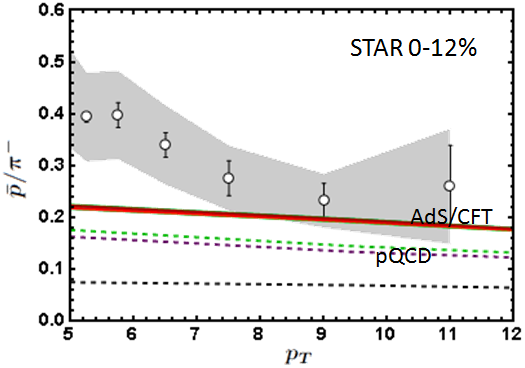}
\end{array}$
\end{center}
\vspace{-.1in}
\caption{\label{figAdS}(Left) A simple energy loss model based on AdS/CFT gives a reasonably good qualitative description of the \pizero \raapt suppression.  Data are from \cite{Adare:2008qa}.  (Right) The AdS/CFT-based energy loss model does a better job of describing the $\bar{p}/\pi^-$ ratio measured by STAR \cite{Abelev:2007ra} than that based on pQCD (although the disagreement between pQCD and the measurement is decreased significantly with the use of modern DSS fragmentation functions \cite{deFlorian:2007aj} instead of calculations \cite{Fries:2003kq} based on the older KKP \cite{Kniehl:2000fe}).}
\end{figure}

\vspace{-.05in}
An energy loss model that correctly describes the relevant physics in heavy ion collisions should describe well all measured observables simultaneously.  We see that this is difficult to do for the current set of pQCD-based energy loss models.  One naturally asks, then, can AdS/CFT-based energy loss models describe more than just non-photonic electron suppression \cite{Horowitz:2010dm,Akamatsu:2008ge}?  Recent derivations \cite{Gubser:2008as,Chesler:2008uy} using AdS/CFT have opened the possibility of describing the suppression pattern of light quarks and gluons in a QGP.  Specifically, the claim is that in a strongly coupled medium an off-shell light quark or gluon thermalizes in a distance $L_\mathrm{therm}\propto(C_R E/T)^{1/3}/T$.  Using this as a basis for a very simple $\theta$-function energy loss model (the particle escapes if its pathlength $L$ is shorter than $L_\mathrm{therm}$ and it disappears if $L>L_\mathrm{therm}$), we can see in the left panel of \fig{figAdS} that, within the rather large theoretical uncertainties associated with these energy loss calculations, the suppression from the AdS/CFT model is in qualitative agreement with the data while simultaneously also in qualitative agreement with the non-photonic electron suppression \cite{Horowitz:2010dm,Akamatsu:2008ge}.  We also see in the right panel of \fig{figAdS} that the AdS result is also in better qualitative agreement with the $\bar{p}/\pi^-$ ratio measured by STAR \cite{Abelev:2007ra}; this better agreement is because the color factor in the AdS/CFT-based model comes to the power 1/3, as opposed to the power 1 in pQCD-based models.  

\vspace{-.05in}
\section{Conclusions}
\vspace{-.05in}
While there are large systematic theoretical uncertainties in pQCD-based energy loss models, describing multiple experimental observables simultaneously seems to provide a means of falsifying these calculations.  In particular the very large very \highpt \pizero \vtwo data presents a significant challenge for pQCD-based models to describe.  Since the WHDG energy loss model does give a very good description of the centrality dependence of the out-of-plane \raacomma, a better treatment of the coupling of energy loss and flow may resolve this \vtwo puzzle.  On the other hand energy loss calculations based on the strong coupling derivations of AdS/CFT seem to give a better qualitative agreement between the non-photonic electron and \pizero suppression as well as ratios such as $\bar{p}\pi^-$.  Future measurements at the LHC, such as the double ratio of charm to bottom \raa shown in \cite{Horowitz:2007su}, might well help demonstrate qualitatively, and independently from hydrodynamical calculations, whether or not the QGP created in heavy ion collisions is a strongly coupled liquid or a weakly coupled plasma.  

\vspace{-.05in}
\section{Acknowledgements}
\vspace{-.05in}
The author wishes to thank Brian Cole, Miklos Gyulassy, Ulrich Heinz, Jiangyong Jia, Thorsten Renk, and Yuri Kovchegov for fruitful discussions.  This work was sponsored in part by the U.S.\ Department of Energy under Grant No.\ DE-SC0004286, and we thank the Institute for Nuclear Theory at the University of Washington for its hospitality and the Department of Energy for partial support during the completion of this work.





\providecommand{\href}[2]{#2}\begingroup\raggedright\endgroup







\end{document}